\documentclass{INTERSPEECH2023}

\usepackage{multirow, graphicx}
\usepackage{amsmath,graphicx,multirow}
\usepackage{mathrsfs}
\usepackage{color}
\usepackage{CJKutf8}
\usepackage{makecell}
\usepackage{cite}

\interspeechcameraready
\title{FunASR: A Fundamental End-to-End Speech Recognition Toolkit}
\name{Zhifu Gao, Zerui Li, Jiaming Wang, Haoneng Luo, Xian Shi, Mengzhe Chen, Yabin Li, Lingyun Zuo, Zhihao Du, Zhangyu Xiao, Shiliang Zhang}
\address{Speech Lab of DAMO Academy, Alibaba Group, China}
\email{\{zhifu.gzf, lzr265946, wangjiaming.wjm, haoneng.lhn, shixian.shi, mengzhe.cmz, wucong.lyb, ailsa.zly, neo.dzh, jiangyu.xzy, sly.zsl\}@alibaba-inc.com}

\begin{document}

\maketitle
 
\begin{abstract}

This paper introduces FunASR\footnote{https://github.com/alibaba-damo-academy/FunASR}, an open-source speech recognition toolkit designed to bridge the gap between academic research and industrial applications. FunASR offers models trained on large-scale industrial corpora and the ability to deploy them in applications. The toolkit's flagship model, Paraformer, is a non-autoregressive end-to-end speech recognition model that has been trained on a manually annotated Mandarin speech recognition dataset that contains 60,000 hours of speech. To improve the performance of Paraformer, we have added timestamp prediction and hotword customization capabilities to the standard Paraformer backbone. In addition, to facilitate model deployment, we have open-sourced a voice activity detection model based on the Feedforward Sequential Memory Network (FSMN-VAD) and a text post-processing punctuation model based on the controllable time-delay Transformer (CT-Transformer), both of which were trained on industrial corpora. These functional modules provide a solid foundation for building high-precision long audio speech recognition services. Compared to other models trained on open datasets, Paraformer demonstrates superior performance.

\end{abstract}
\noindent\textbf{Index Terms}: FunASR, Paraformer, Speech Recognition, FSMN-VAD, CT-Transformer

\section{Introduction}

Over the past few years, the performance of end-to-end~(E2E) models has surpassed that of conventional hybrid systems on automatic speech recognition~(ASR) tasks.
There are three popular E2E approaches: connectionist temporal classification (CTC)~\cite{graves2006connectionist}, recurrent neural network transducer (RNN-T)~\cite{graves2013speech} and attention based encoder-decoder (AED)~\cite{chan2016listen,vaswani2017attention}. 
Of these, AED models have dominated seq2seq modeling for ASR, due to their superior recognition accuracy~\cite{vaswani2017attention,gulati2020conformer,gao2020san,zhang2020streaming,gao2020universal,radfar22_interspeech,sklyar22_interspeech,do22_interspeech,lee22g_interspeech,zhao22f_interspeech}. 
Open-source toolkits including ESPNET~\cite{watanabe2018espnet}, WeNet~\cite{yao2021wenet}, PaddleSpeech~\cite{zhang2022paddlespeech} and K2~\cite{kang2022fast} et al., have been developed to facilitate research in end-to-end speech recognition. These open-source tools have played a great role in reducing the difficulty of building an end-to-end speech recognition system.

In this work, we introduce FunASR, a new open source speech recognition toolkit designed to bridge the gap between academic research and industrial applications. FunASR builds upon previous works and provides several unique features:
\begin{enumerate}
    \item \emph{Modelsope}: FunASR provides a comprehensive range of pre-trained models based on industrial data. The flagship model, Paraformer~\cite{gao2022paraformer}, is a non-autoregressive end-to-end speech recognition model that has been trained on a manually annotated Mandarin speech recognition dataset that contains 60,000 hours of speech.  Compared with Conformer~\cite{gulati2020conformer} and RNN-T~\cite{graves2013speech} supported by mainstream open source frameworks, Paraformer offers comparable performance while being more efficient. 
\begin{figure}
    \centering
    \includegraphics[width=0.9\linewidth]{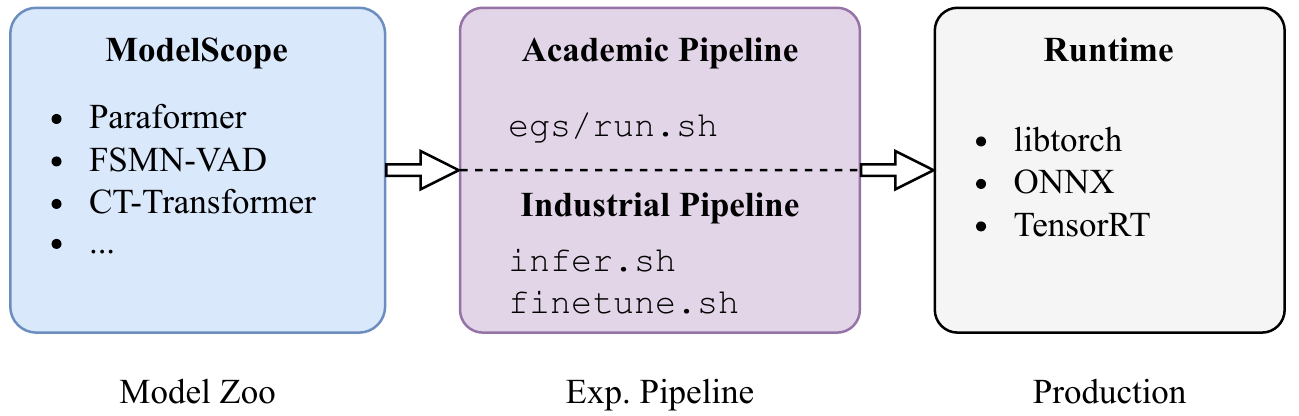}
    \caption{Overview of FunASR design.}
    \label{fig:overview}
    \vspace{-8mm}
\end{figure}

    \item \emph{Training \& Finetuning}:  FunASR is a comprehensive toolkit that offers a range of example recipes to train end-to-end speech recognition models from scratch, including Transformer, Conformer, and Paraformer models for datasets like  AISHELL~\cite{bu2017aishell,du2018aishell}, WenetSpeech~\cite{zhang2022wenetspeech} and LibriSpeech~\cite{panayotov2015librispeech}. Additionally, FunASR provides a convenient finetuning script that allows users to quickly fine-tune a pre-trained model from the ModelScope on a small amount of domain data, resulting in high-performance recognition models. This feature is particularly beneficial for academic researchers and developers who may have limited access to data and computing power required to train models from scratch.

    \item \emph{Speech Recognition Services}: FunASR enables users to build speech recognition services that can be deployed on real-applications.
    To facilitate model deployment, we have also released a voice activity detection model based on the Feedforward Sequential Memory Network (FSMN-VAD)~\cite{zhang2018deep} and a text post-processing punctuation model based on the controllable time-delay Transformer (CT-Transformer)~\cite{chen2020controllable}, both of which were trained on industrial corpora. To improve the performance of Paraformer, we have added timestamp prediction and hotword customization capabilities to the standard Paraformer backbone. Additionally, FunASR includes an inference engine that supports CPU and GPU inference through ONNX, libtorch, and TensorRT. These functional modules simplify the process of building high-precision, long audio speech recognition services using FunASR.
\end{enumerate}
Overall, FunASR is a powerful speech recognition toolkit that offers unique features not found in other open source tools. We believe that our contributions will help to further advance the field of speech recognition and enable more researchers and developers to apply these techniques to real-world applications.
It should be noted that, this paper only reports the experiments on Mandarin corpora, due to the limitation of the number of pages. In fact, FunASR supports many types of languages, including English, French, German, Spanish, Russian, Japanese, Korean, etc (more details could be found in model zoo).

\begin{figure*}[h]
    \vspace{-3mm}
    \centering
    \includegraphics[width=0.9\textwidth]{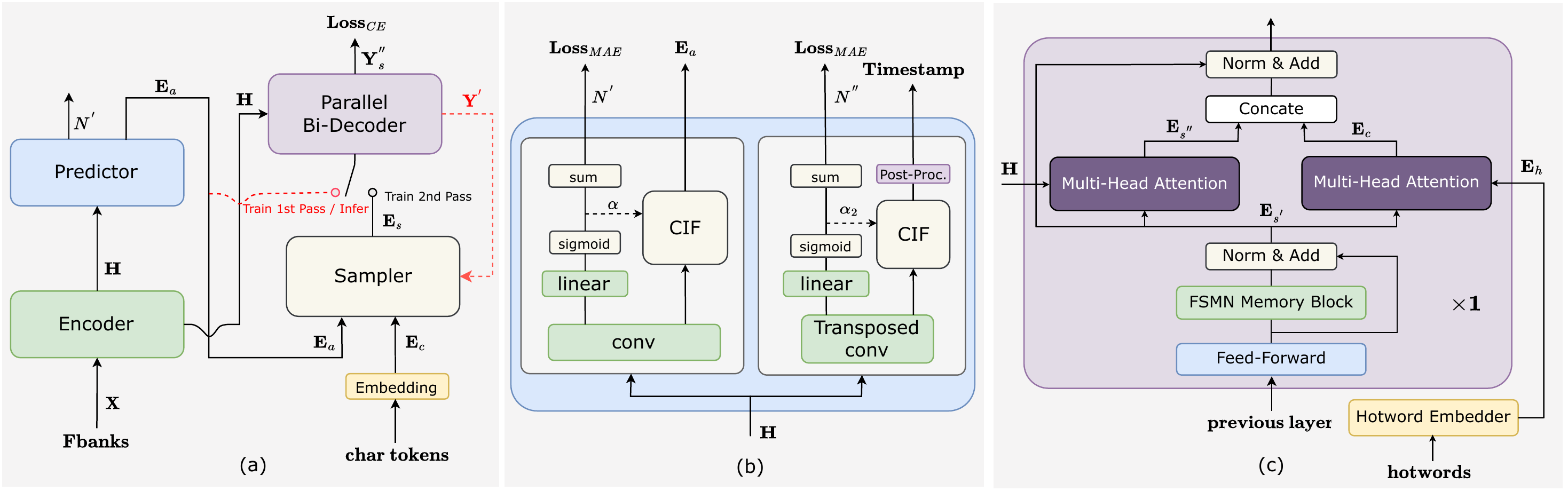}
    \caption{Illustrations of the Paraformer related architectures. (a) Paraformer; (b) Advanced timestamp prediction; (c) Contextual decoder layer for hotword customization.}\label{arch}
    \vspace{-6mm}
\end{figure*}

\section{Overview of FunASR}

The overall framework of FunASR is illustrated in Fig.~\ref{fig:overview}. ModelScope manages the models utilized in FunASR and hosts critical ones such as Paraformer, FSMN-VAD, and CT-Transformer.

Users of FunASR can easily perform experiments using its Pytorch-based pipelines, which are categorized as either academic or industrial pipelines. The academic pipeline, denoted by $run.sh$, enables users to train models from scratch.
The $run.sh$ script follows the recipe style of ESPNET and includes stages for data preparation (stage 0), feature extraction (stage 1), dictionary generation (stage 2), model training (stage 3 and 4), and model inference and scoring (stage 5). In contrast, the industrial pipeline offers two separate scripts: $infer.sh$ for inference and $finetune.sh$ for fine-tuning. These pipelines are easy to use, with users only needing to specify the model name and dataset.

FunASR also provides an easy-to-use runtime for deploying models in applications. To support various hardware platforms such as CPU, GPU, Android, and iOS, we offer different runtime backends including Libtorch, ONNX, and TensorRT. In addition, we utilize AMP quantization~\cite{gao21f_interspeech} to accelerate the inference runtime and ensure optimal performance. With these features, FunASR makes it easy to deploy and use speech recognition models in a wide range of applications.

\section{Main Modules of FunASR}

\subsection{Paraformer}

To begin, let us provide a brief overview of Paraformer~\cite{gao2022paraformer}, a model that we have previously proposed, depicted in Fig.~\ref{arch}(a). Paraformer is a single-step non-autoregressive (NAR) model that incorporates a glancing language model-based sampler module to enhance the NAR decoder's ability to capture token inter-dependencies.

The Paraformer consists of two core modules: the predictor and the sampler. The predictor module is used to generate acoustic embeddings, which capture the information from the input speech signals. 
\
During training, the sampler module incorporates target embeddings by randomly substituting tokens into the acoustic embeddings to generate semantic embeddings. This approach allows the model to capture the interdependence between different tokens and improves the overall performance of the model. However, during inference, the sampler module is inactive, and the acoustic embeddings are used to output the final prediction over only a single pass. This approach ensures faster inference times and lower latency.

In order to further enhance the performance of Paraformer, this paper proposes modifications including timestamp prediction and hotword customization. In addition, the loss function used in~\cite{gao2022paraformer} has been updated by removing the MWER loss, which was found to contribute little to performance gains. An additional CE loss is now used in the first pass decoder to reduce the discrepancy between training and inference. The next subsection will provide detailed explanations.

\subsection{Timestamp Predictor}

Accurate timestamp prediction is a crucial function of ASR systems. However, conventional industrial ASR systems require an extra hybrid model to conduct force-alignment (FA) for timestamp prediction (TP), leading to increased computation and time costs. FunASR provides an end-to-end ASR model that achieves accurate timestamp prediction by redesigning the structure of the Paraformer predictor, as depicted in Fig.\ref{arch}(b). We introduce a transposed convolution layer and LSTM layer to upsample the encoder output, and timestamps are generated by post-processing CIF~\cite{dong2020cif} weights $\alpha_2$. We treat the frames between two fireplaces as the duration of the former tokens and mark out the silence parts according to $\alpha_2$. In addition, FunASR also releases a \textit{force-alignment-like} model named TP-Aligner, which includes an encoder of smaller size and a timestamp predictor. It takes speech and corresponding transcription as input to generate timestamps.

\begin{table}[h]
\vspace{-1.5mm}
\centering
\caption{Evaluation of timestamp prediction.}\label{tabel-ts}
\vspace{-1.5mm}
\resizebox{\linewidth}{!}{
\begin{tabular}[h]{lll}

\hline
Data                             & System                & AAS~(ms) \\ \hline
\multirow{2}{*}{AISHELL}       & Force-alignment       & 80.1          \\
                                 & Paraformer-TP        & 71.0          \\ \hline
\multirow{3}{*}{Industrial Data} & Force-alignment       & 60.3          \\
                                 & Paraformer-large-TP & 65.3          \\
                                 & TP-Aligner         & 69.3          \\ \hline
\end{tabular}}
\vspace{-3mm}
\end{table}

We conducted experiments on AISHELL and 60,000-hour industrial data to evaluate the quality of timestamp prediction. The evaluation metrics used for measuring timestamp quality is the accumulated average shift (AAS)\cite{shi2023achieving}. We used a test set of 5,549 utterances with manually marked timestamps to compare the timestamp prediction performance of the provided models with FA systems trained with Kaldi\cite{povey2011kaldi}.
The results show that Paraformer-TP outperforms the FA system on AISHELL. In industrial experiments, we found that the proposed timestamp prediction method is comparable to the hybrid FA system in terms of timestamp accuracy (with a gap of less than 10ms). Moreover, the one-pass solution is valuable for commercial usage as it helps in reducing computation and time overhead.

\subsection{Hotword Customization}
Contextual Paraformer offers the ability to customize hotwords by utilizing named entities, which enhances incentives and improves the recall and accuracy. Two additional modules have been added to the basic Paraformer model - a hotword embedder and a multi-head attention in the last layer of the decoder, depicted in  Fig.~\ref{arch}(c). 

We utilize hotwords, denoted as $\bm{w} = \bm{w{_1}},...,\bm{w{_n}}$, as input to our hotword embedder~\cite{pundak2018deep}. The hotword embedder consists of an embedding layer and LSTM layer, which takes the context hotwords as input and generates an embedding, denoted as ${\bm{E_{h}}}$, by using the last state of the LSTM. Specifically, the hotwords are first fed to the hotword embedder, which produces a sequence of hidden states. We then use the last hidden state as the embedding of the hotwords, capturing the contextual information of the input sequence.

To capture the relationship between the hotword embedding ${\bm{E_{h}}}$ and the output of the last layer of the FSMN memory block ${\bm{E_{s^{'}}}}$, we employ a multi-head attention module. Then, we concate the ${\bm{E_{s^{'}}}}$ and contextual attention ${\bm{E_{c}}}$. This operation is formalized in Equation~\ref{eq1}:

\begin{equation}
\begin{gathered}
E_{c} = \text {MultiHeadAttention}(E_{s^{'}}W_{c}^{Q}, E_{h}W_{c}^{K}, E_{h}W_{c}^{V}), \\
E_{s^{''}} = \text {MultiHeadAttention}(E_{s^{'}}W_{s}^{Q}, HW_{s}^{K}, HW_{s}^{V}), \\
O = \text {Conv1d} ([E_{s^{''}};E_{c}])
\end{gathered}
\label{eq1}
\end{equation}

We use a one-dimensional convolutional layer (${Conv1d}$) to reduce its dimensionality to match that of the hidden state ${\bm{E_{s^{'}}}}$, which serves as the input of the subsequent layer. It's worth noting that apart from this modification, the other processes of our Contextual Paraformer are the same as those of the standard Paraformer. 

\begin{table}[h]
\vspace{-1mm}
    \centering
    \caption{The test sets used in this customization task.}
    \vspace{-2mm}
    \begin{tabular}{ccc}
        \midrule
        Dataset & Utts & Named Entities \\
        \midrule
        AI domain & 486 & 204 \\
        Common domain & 1308 & 231 \\
        \midrule
    \end{tabular}
    \label{tab:customization_testset}
    \vspace{-3mm}
\end{table}

During the training, the hotwords are randomly generated from target in each training batch. As for inference, we can specify hotwords  by providing a list of named entities to the model.

\begin{table}[h]
\vspace{-1mm}
    \centering
    \caption{Evaluation of hotword customization}
    \vspace{-2mm}
    \resizebox{\linewidth}{!}{
    \begin{tabular}[h]{cccccc}
        \midrule
        Dataset & Hotwords & CER & R & P & F$_1$ \\
        \midrule
        \multirow{2}{*}{AISHELL hotword subtest} &w/o & 10.01 & 16 & 100 & 27 \\
        &w/ & 4.55 & 74 & 100 & 85 \\
        \hline
        \multirow{2}{*}{Industrial AI domian} &w/o & 7.96 & 70 & 98 & 82 \\
        &w/  & 6.31 & 89 & 98 & 93 \\
        \hline
        \multirow{2}{*}{Industrial common domian} &w/o  & 9.47 & 67 & 100 & 80 \\
        &w/  & 8.75 & 80 & 98 & 88 \\
        \midrule
    \end{tabular}}
    \label{tab:hotword}
    \vspace{-3mm}
\end{table}

To evaluate the hotword customization effect of Contextual Paraformer, we created a hotword testset by sampling 235 audio clips containing entity words from the AISHELL testset, which included 187 named entities. The dataset has been uploaded to the ModelScope and the test recipe has been opened to FunASR. Additionally, we expanded our experiments to include the AI domain and Common domain of industrial tasks, as presented in Tab.~\ref{tab:customization_testset}.

\begin{figure}
    \centering
    \includegraphics[width=0.7\linewidth]{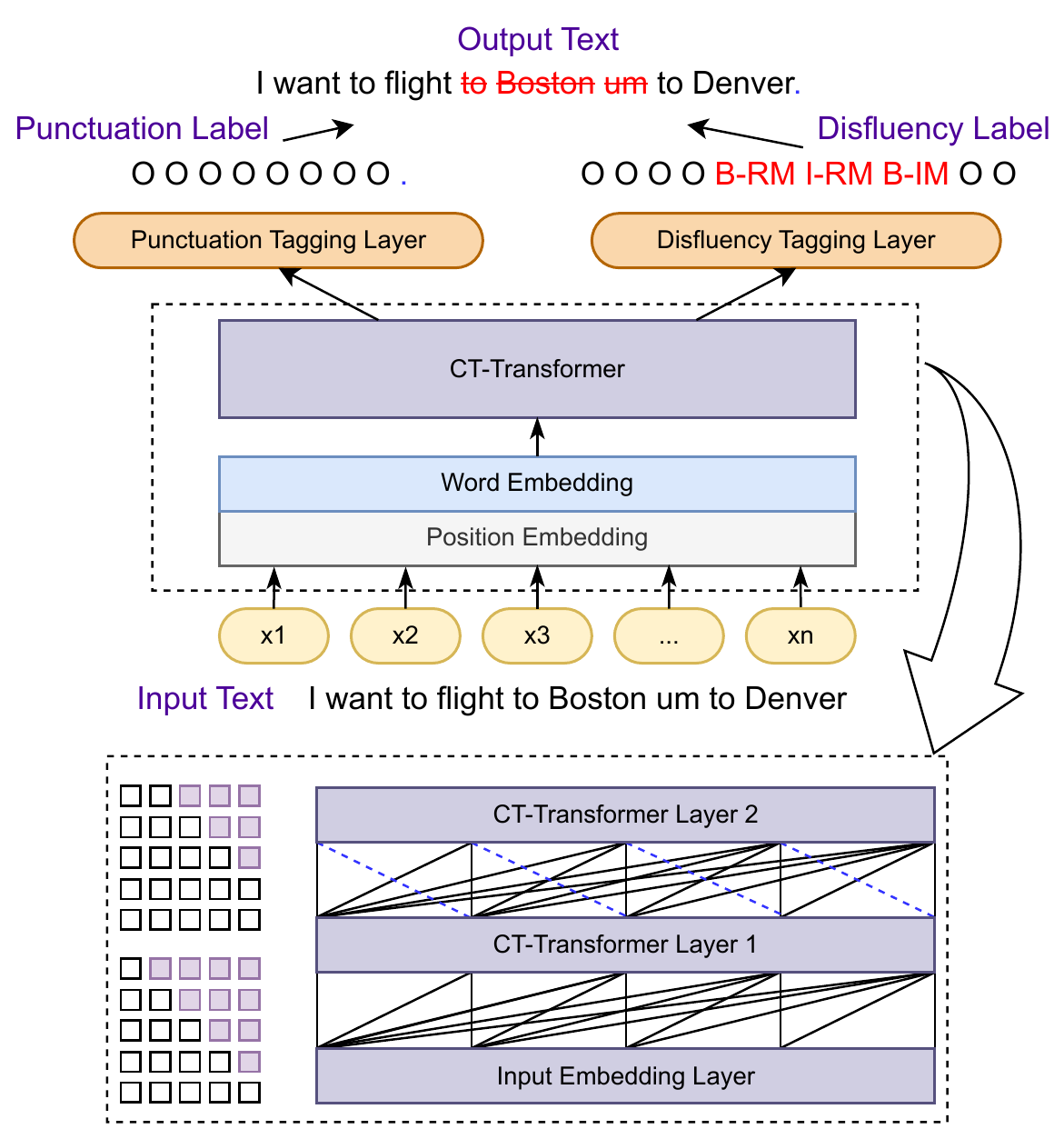}
    \caption{Illustration of CT-Transformer architecture.}
    \label{fig:CT-Transformer}
    \vspace{-3mm}
\end{figure}

Tab.~\ref{tab:hotword} presents the experimental results of our study on the impact of hotwords on the performance of Contextual Paraformer. We employed CER and F1-score as evaluation metrics for the customization task. Our results indicate an impressive improvement of approximately 58\% in F1-score on the AISHELL-1 named entity subtest. Moreover, we achieved an average of 10\% improvement on industrial customization tasks.

\subsection{Voice Activity Detection}

Voice activity detection (VAD) plays a important role in speech recognition systems by detecting the beginning and end of effective speech. FunASR provides an efficient VAD model based on the FSMN structure~\cite{zhang2018deep}. To improve model discrimination, we use monophones as modeling units, given the relatively rich speech information. During inference, the VAD system requires post-processing for improved robustness, including operations such as threshold settings and sliding windows.

\begin{table}[h]
    \vspace{-1mm}
    \centering
    \caption{Evaluation of VAD on continuous utterances.}
    \vspace{-2mm}
    \begin{tabular}{ccccc}
        \midrule
        Dataset & VAD & CER & Speech$\%$ \\
        \hline
        \multirow{2}{*}{Meeting domain} &w/o & 2.42 & 1 \\
        &w/ & 2.31 & 0.92 \\
        \hline
        \multirow{2}{*}{Video domain} &w/o & 27.87 & 1 \\
        &w/ & 25.34 & 0.78 \\
        \midrule
    \end{tabular}
    \label{tab:vad_res}
    \vspace{-3mm}
\end{table}

\begin{table*}[h]
    \vspace{-3mm}
    \centering
    \caption{Comparison of CERs on AISHELL, AISHELL-2 and WenetSpeech with open source speech toolkits. RTF is evaluated with batchsize 1.($\dag$ : The number of the model parameters is 110M) }
    \vspace{-2mm}
    \begin{tabular}{c|ccccccc}
        \midrule
        Model & \#Parameters & AR/NAR & LM  & \makecell[c]{AISHELL\\ test} & \makecell[c]{AISHELL-2 \\ test\_ios} & \makecell[c]{WenetSpeech \\ test\_meeting} & RTF \\
        \midrule
        ESPNET Conformer~\cite{watanabe2018espnet} & 46.25M & AR &w/ & 4.60 & 5.70 & 15.90 $\dag$ & - \\
        WeNet Conformer-U2++~\cite{zhang2022wenet} & 47.30M & AR &w/ & 4.40 & 5.35 & 17.34 $\dag$ & -\\
        PaddleSpeech DeepSpeech2~\cite{zhang2022paddlespeech} & 58.4M & NAR &w/& 6.40 & - & - &- \\
        K2 Transducer~\cite{kang2022fast} & 80M & AR &w/o& 5.05 & 5.56 & 14.44 $\dag$ & - \\
        \hline
        \multirow{2}{*}{Conformer} & \multirow{2}{*}{46.25M} & \multirow{2}{*}{AR} &w/& 4.65 & 5.35 & 15.21 & 1.4300 \\
         &  &  &w/o& 5.21 & 5.83  & - & 0.2100 \\
        Paraformer & 46.3M & NAR &w/o& 4.95 & 5.73 & - & 0.0168\\
        Paraformer-large & 220M & NAR &w/o& 1.95 & 2.85& 6.97 & 0.0251 \\
        \midrule
    \end{tabular}
    \label{tab:paraformer_res}
    \vspace{-5mm}
\end{table*}
%

The evaluation of VAD is presented in detail in Table ~\ref{tab:vad_res}. The test set consists of manually annotated data from two domains: 2 hours of meeting data and 4 hours of video data. We report the Character Error Rate (CER) and the percentage of utterances sent to ASR inference for recognition. The results demonstrate that the VAD effectively filters out invalid voice, allowing the recognition system to focus on effective speech and leading to a significant CER improvement.

\subsection{Text Postprocessing}
Text postprocessing is a critical step in generating readable ASR transcripts, which involves adding punctuation marks and removing speech disfluencies. FunASR includes a CT-Transformer model that performs both tasks in real-time, as described in~\cite{chen2020controllable}. The model's overall framework is presented in Fig.~\ref{fig:CT-Transformer}. To meet real-time constraints, the model allows partial outputs to be frozen with controllable time delay. A fast decoding strategy is utilized to minimize latency while maintaining competitive performance. Moreover, to reduce computational complexity, the strategy dynamically discards a history that is too long based on already predicted punctuation marks.

\begin{table}[h]
\vspace{-1mm}
\caption{The results of text postprocessing}
\vspace{-5mm}
\begin{center}
\scalebox{0.78}
{
\begin{tabular}{c c c c c c c c}
\hline
\multirow{2}{*}{\textbf{Model}} & 
\multicolumn{3}{c}{\textbf{Punctuation}} & 
\multicolumn{3}{c}{\textbf{Disfluency}} &
\multirow{2}{*}{\textbf{Inference Time}}\\
& P & R & F$_1$ & P & R & F$_1$ \\
\hline
BLSTM  & 60.2 & 48.8 & 53.9  & 84.1 & 57.0 & 67.9 & 1112.9s ($\times1.0$)\\
Full-Trans. & 62.1 & 55.9 & 58.8 & 83.1 & 61.2 & 70.5 & 676.7s ($\times1.6$)\\
CT-Trans. & 62.7 & 55.3 & 58.8 & 82.4 & 61.5 & 70.5 & 585.8s ($\times1.9$)\\
\hline
\end{tabular}
}
\end{center}
\label{tab:punc_disf_res}
\vspace{-6mm}
\end{table}

We evaluated the punctuation prediction and disfluency detection on an in-house Chinese dataset that consists of about 24K spoken utterances with punctuation and disfluency annotations. Tab.~\ref{tab:punc_disf_res} shows the precision (P), recall (R), and F$_1$-score (F$_1$) of the models. The results indicate that the CT-Transformer achieves competitive F$_1$ with a faster inference speed.

\section{Experiments}

\subsection{Evaluation ASR}

In our experiments, we evaluate the performance of our models on the AISHELL, AISHELL-2, and WenetSpeech datasets and present the results in detail in Table~\ref{tab:paraformer_res}. When compared to other open source toolkits, FunASR achieves comparable results to the baseline Conformer model. To ensure a fair comparison with Paraformer, we remove the language models (LM) and joint-CTC decoding of Conformer. The results show that Paraformer slightly outperforms Conformer without LM in terms of recognition accuracy. We also evaluate the computational efficiency of the models in terms of inference speed by RTF on GPU (V100) with a batch size of 1. The Paraformer model achieves a 12x speedup in inference, which is a significant advantage over autoregressive models. Even when the number of parameters in the Paraformer-large model is increased from 46M to 220M, the RTF still outperforms the AR model with 46M parameters.

FunASR provides a pre-trained Paraformer-large model that is specifically trained on a 60,000 hour Mandarin speech recognition corpus used in industry. The performance of Paraformer-large is impressive, as shown in Table~\ref{tab:paraformer_res}. It achieves low CERs of 1.95\%, 2.85\%, and 6.97\% on the AISHELL test, AISHELL-2 test\_ios, and WenetSpeech test\_meeting task, respectively. These results demonstrate the importance of using large-scale speech corpora in improving the performance of ASR systems.

Moreover, FunASR offers customization capabilities for our pretrain model, allowing it to be finetuned on domain-specific data. The results of using the AISHELL and industrial domain data (200h) to finetune our Paraformer-large model are presented in Table~\ref{tab:finetune_res}. Our model achieved relative improvements of 7.4\% and 8.7\% on the AISHELL dev and test tasks, respectively. Additionally, the experiments in the logistics field have demonstrated that finetuning the Paraformer-large model results in a significant improvement in recognition and domain keywords recall. The short testset consists of audio files that are 5.77s long on average, with a total duration of 2.41 hours. The long testset is 4.95 hours long and contains audio with an average length of 162.0s. On average, we observed an increase in domain keywords recall from 76.7\% to 96.8\%, and a decrease in CER from 11.25\% to 10.10\%. The experiments have demonstrated that the pretrained model can be finetuned with your corpus to achieve significant improvements in the relevant field.

\begin{table}[h]
    \vspace{-1mm}
    \centering
    \caption{Evaluation of funetuning}
    \vspace{-2mm}
    \resizebox{\linewidth}{!}{
    \begin{tabular}{c|c|c|c}
        \midrule
        Model & Dataset &{Pretrain(CER\%/Recall\%)} & {Finetune(CER\%/Recall\%)}\\
        \midrule
        \multirow{2}{*}{AISHELL} & dev & 1.75 / - & 1.62  / - \\
         & test & 1.95 / - & 1.78 / - \\
        \hline
        \multirow{2}{*}{\makecell[c]{Logistics \\ domain}} & short & 9.4 / 79.4 & 7.6  / 97.6 \\
         & long & 13.1 / 74.0 & 12.6 / 96.0 \\
        \midrule
    \end{tabular}}
    \label{tab:finetune_res}
    \vspace{-6mm}
\end{table}

\subsection{Runtime Benchmark}

This section evaluates the runtime performance of Paraformer-large in terms of both CER and RTF on an Intel(R) Xeon(R) CPU E5-8269CY @ 2.5GHz with one thread core. We evaluate the performance of Paraformer-large with two runtime backends, as shown in Table~\ref{tab:runtime}. To optimize performance, FunASR adopts the automatic mixed precision quantization (AMP) method proposed in~\cite{gao21f_interspeech,zhu2021disc}.
The results demonstrate that AMP quantization improves inference speed by 40\% without any significant reduction in recognition accuracy.

\begin{table}[h]
    \vspace{-1mm}
    \centering
    \caption{Runtime benchmark.}
    \vspace{-3mm}
    \begin{tabular}{c|cccc}
        \midrule
        \multirow{2}{*}{Quantization} & \multicolumn{2}{c}{Libtorch} & \multicolumn{2}{c}{Onnx} \\
         & Float32 & Int8 & Float32 & Int8 \\
        \midrule
        RTF & 0.1026 & 0.0597 & 0.0778 & 0.0446 \\
        CER & 1.95 & 1.95 & 1.95 & 1.95 \\
        \midrule
    \end{tabular}
    \label{tab:runtime}
    \vspace{-8mm}
\end{table}

\section{Conclusions}

This paper presents FunASR, a system designed to bridge the gap between academic research and industrial applications in speech recognition. FunASR provides access to models trained on large-scale industrial corpora, as well as the ability to easily deploy them in real-world applications. We make a wide range of industrial models available, including  Paraformer-large model, as well as FSMN-VAD and CT-Transformer models, etc.
By making these models openly available, FunASR enables researchers to easily deploy them in real-world scenarios.

\vfill\pagebreak

\bibliographystyle{IEEEtran}
\bibliography{mybib}

\end{document}